\pdfoutput=1

\documentclass[11pt,twoside,a4paper,cmspaper,final,collab]{cms-tdr}
\def\svnVersion{274665}\def\cmsCernNo{2013-037}\def\cmsCernDate{\today}\def\cmsMessage{Published in the Journal of High Energy Physics as \href{http://dx.doi.org/10.1007/JHEP01(2015)096}{\doi{10.1007/JHEP01(2015)096.}}}

\begin{document}\cmsNoteHeader{EXO-12-034}

\hyphenation{had-ron-i-za-tion}
\hyphenation{cal-or-i-me-ter}
\hyphenation{de-vices}
\RCS$Revision: 274636 $
\RCS$HeadURL: svn+ssh://svn.cern.ch/reps/tdr2/papers/EXO-12-034/trunk/EXO-12-034.tex $
\RCS$Id: EXO-12-034.tex 274636 2015-01-22 16:37:18Z wulsin $
\newlength\cmsFigWidth
\ifthenelse{\boolean{cms@external}}{\setlength\cmsFigWidth{0.48\textwidth}}{\setlength\cmsFigWidth{0.6\textwidth}}
\ifthenelse{\boolean{cms@external}}{\providecommand{\cmsLeft}{top}}{\providecommand{\cmsLeft}{left}}
\ifthenelse{\boolean{cms@external}}{\providecommand{\cmsRight}{bottom}}{\providecommand{\cmsRight}{right}}
\newcommand{\calotot}{\ensuremath{E_\text{calo}}\xspace}
\newcommand{\calototCut}{\ensuremath{E_\text{calo}<10\GeV}\xspace}
\newcommand{\candtrk}{{candidate track}\xspace}
\newcommand{\Chipm}{\ensuremath{\chipm_1}\xspace}
\newcommand{\Chiz}{\ensuremath{\chiz_1}\xspace}
\newcommand{\ctau}{\ensuremath{c \tau}\xspace}
\newcommand{\dishtrk}{{disappearing-track}\xspace}
\newcommand{\distrk}{{disappearing track}\xspace}
\newcommand{\Ecal}{ECAL\xspace}
\newcommand{\Hcal}{HCAL\xspace}
\newcommand{\Nmissin}{\ensuremath{N_\text{inner}}\xspace}
\newcommand{\Nmissmid}{\ensuremath{N_\text{mid}}\xspace}
\newcommand{\Nmissout}{\ensuremath{N_\text{outer}}\xspace}
\newcommand{\pp}{\ensuremath{\Pp\Pp}\xspace}
\newcommand{\qq}{\ensuremath{\Pq\Paq}\xspace}
\newcommand{\qqp}{\ensuremath{\Pq\Paq'}\xspace}
\newcommand{\Wjets}{\ensuremath{\PW\to\ell\nu}+jets\xspace}
\newcommand{\Zee}{\ensuremath{\cPZ\to\EE}\xspace}
\newcommand{\Zll}{\ensuremath{\cPZ\to\ell\ell}\xspace}
\newcommand{\Zmumu}{\ensuremath{\cPZ\to\MM}\xspace}
\newcommand{\Ztautau}{\ensuremath{\cPZ\to\TT}\xspace}
\newcommand{\tauh}{\ensuremath{\Pgt_\mathrm{h}}\xspace}

\cmsNoteHeader{EXO-12-034}
\title{Search for disappearing tracks in proton-proton collisions at $\sqrt{s} = 8$\TeV}

\date{\today}

\abstract{
A search is presented for long-lived charged particles that decay
  within the
  CMS detector and produce the signature of a disappearing track.
  Disappearing tracks are identified as those with little or no
  associated calorimeter energy deposits and with
  missing hits in the outer layers of the tracker.
  The search uses proton-proton collision data recorded at $\sqrt{s}=8\TeV$ that
  corresponds to an integrated luminosity of 19.5\fbinv.
  The results of the search are interpreted in the context of the
  anomaly-mediated supersymmetry breaking (AMSB) model.
The number of observed events is in agreement with the background
expectation, and limits are set on the cross section of direct electroweak
chargino production in terms of the chargino mass and mean proper lifetime.
At 95\% confidence level, AMSB models with a chargino mass less than
260\GeV, corresponding to a mean proper lifetime of  0.2\unit{ns},  are excluded.
}

\hypersetup{%
pdfauthor={CMS Collaboration},%
pdftitle={Search for disappearing tracks in proton-proton collisions at sqrt(s) = 8 TeV},%
pdfsubject={CMS},%
pdfkeywords={CMS, physics, software, computing}}

\maketitle 

\section{Introduction}
\label{sec:intro}

We present a search for long-lived charged particles that
decay
within the tracker volume and produce the signature of a
\textit{\distrk}.  A disappearing track can be produced in beyond the standard model (BSM) scenarios
by a charged particle whose decay products are undetected. This occurs because the decay products are
either too low in momentum to be reconstructed or neutral (and weakly
interacting) such that they do not interact with the tracker material
or deposit significant energy in the calorimeters.

There are many BSM scenarios that produce particles that manifest themselves as disappearing tracks
~\cite{PhysRevD.85.095011, SpreadSUSY, MiniSplitSUSY, UnnaturalSUSY, Ellis2012}.  One example is anomaly-mediated supersymmetry breaking (AMSB) \cite{Giudice1998, Randall1999}, which predicts a particle mass
spectrum that has a small mass splitting between the lightest
chargino (\Chipm) and the lightest neutralino (\Chiz). The chargino can then decay to a neutralino and a pion,
$\Chipm \to \Chiz \Pgp^\pm$. The phase space for this decay is limited by the small
chargino-neutralino mass splitting. As a consequence, the chargino has
a significant lifetime,
and the daughter pion has momentum of ${\approx}100\MeV$, typically
too low for its track to be reconstructed.  For charginos that decay inside the tracker volume, this results in a disappearing track.  We benchmark our
search in terms of its sensitivity to the chargino mass and
chargino-neutralino mass splitting (or equivalently, the chargino mean proper lifetime, $\tau$)
in AMSB.
Constraints are also placed on the chargino mass and mean proper lifetime for
direct electroweak chargino-chargino and chargino-neutralino production.

Previous CMS analyses have searched for long-lived charged particles based
on the signature of anomalous ionization
energy loss~\cite{EXO-11-074, HSCP2011, HSCP2012}, but
none has targeted a disappearing track signature.
A search for disappearing tracks conducted by the ATLAS Collaboration
excludes at 95\% confidence level (CL) a chargino in AMSB scenarios
with mass less than 270\GeV and mean proper lifetime of approximately 0.2
ns~\cite{ATLASDisapp2}.

\section{Detector description and event reconstruction }
\label{sec:detector}

The central feature of the CMS apparatus is a superconducting solenoid
of 6\unit{m} internal diameter.  Within the superconducting solenoid
volume are a  silicon pixel and strip tracker, a lead tungstate
crystal electromagnetic calorimeter (ECAL), and a brass and scintillator
hadron calorimeter (HCAL), each  composed of a barrel and two endcap
sections. Extensive forward calorimetry
complements the coverage provided by the barrel and endcap detectors.
The ECAL consists of 75\,848
crystals that provide coverage in pseudorapidity $\abs{ \eta }< 1.479$
in the barrel region and $1.479 <\abs{ \eta } < 3.0$ in the two
endcap regions.
Muons are measured in gas-ionization detectors embedded in the steel
flux-return yoke outside the solenoid. They are measured
in the pseudorapidity range $\abs{\eta}< 2.4$,
with detection planes made using three technologies: drift tubes,
cathode strip chambers, and resistive plate chambers.
Muons are identified as a track in the central tracker
consistent with either a track or several hits in the
muon system.

The silicon tracker measures ionization energy deposits (``hits'')
from charged particles within the
pseudorapidity range $\abs{\eta}< 2.5$. It consists of 1440 silicon
pixel and 15\,148 silicon strip detector modules and is located in the
3.8\unit{T} field of the superconducting solenoid.  The pixel detector
has three barrel layers and two endcap disks, and the strip tracker
has ten barrel layers and three small plus nine large endcap disks.
Isolated particles with transverse momentum \pt = 100\GeV emitted in the range $\abs{\eta} < 1.4$ have
track resolutions of 2.8\% in \pt and 10 (30)\mum in the transverse
(longitudinal) impact parameter \cite{TRK-11-001}.

The particle-flow (PF) event reconstruction consists in reconstructing and identifying each single particle with an optimized combination of all subdetector information \cite{CMS-PAS-PFT-10-001,  CMS-PAS-PFT-09-001}. The energy of
photons is obtained directly
from the ECAL measurement, corrected for zero-suppression effects. The
energy of electrons is determined from a combination of the track
momentum at the main interaction vertex, the corresponding ECAL
cluster energy, and the energy sum of all bremsstrahlung photons
attached to the track. The energy of muons is taken from the
corresponding track momentum. The energy of charged hadrons is
determined from a combination of the track momentum and the
corresponding ECAL and HCAL energies, corrected for zero-suppression
effects and for the response function of the calorimeters to hadronic
showers. Finally, the energy of neutral hadrons is obtained from the
corresponding corrected ECAL and HCAL energies.

Particles are clustered into jets using the anti-\kt
algorithm~\cite{Cacciari:2008gp} with a
distance parameter of 0.5.  Jet momentum is determined from the
vectorial sum of all particle
momenta in the jet, and is found from simulation to be within 5\% to
10\% of the true momentum over the whole \pt spectrum and detector
acceptance. An offset correction is applied to take into account the
extra energy clustered in jets due to additional proton-proton (pp)
interactions within the same bunch crossing. Jet energy corrections
are derived from the simulation, and are confirmed using in situ
measurements of the energy balance of dijet and photon+jet
events.

The missing transverse
energy \ETslash is defined as
the magnitude of the vector sum of the \pt of all
PF candidates reconstructed in the event.
A more detailed description of the CMS apparatus and event reconstruction, together with a
definition of the coordinate system used and the relevant kinematic
variables, can be found in Ref.~\cite{Chatrchyan:2008zzk}.

\section{Data samples and simulation}
\label{sec:dataset}

The search is performed with $\sqrt{s}= 8$\TeV  pp collision data
recorded in 2012 with the CMS detector at the CERN LHC. The data correspond to an
integrated luminosity of 19.5\fbinv.
A BSM particle that produces a disappearing track would not be identified as
a jet or a particle by the PF algorithm because the track
is not matched to any activity in the calorimeter or muon
systems.
To record such particles with the available triggers, we require one or more
initial-state-radiation
(ISR) jets, against which the BSM particles recoil.
 As a result, the \ETslash is approximately equal to the \pt of the BSM particles, and likewise to the \pt of the ISR jets. To maximize efficiency for the BSM signal, events used for the search are collected with the union of two triggers that had the lowest \ETslash thresholds available during the data taking period.  The first requires $\ETslash > 120\GeV$, where the \ETslash is calculated using the calorimeter information only.  The second trigger requires \ETslash larger than either 95 or 105\GeV, depending on the run period, where \ETslash is reconstructed with the PF algorithm and excludes muons from the calculation.  Additionally, the second trigger requires at least one jet with $\pt > 80\GeV$ within $\abs{\eta} < 2.6$.  The use of alternative \ETslash calculations in these triggers is incidental; the \ETslash thresholds set for these formulations simply happen to be such that these triggers yield the highest BSM signal efficiency.

Events collected with these triggers are required to pass a set of
basic selection criteria. These requirements reduce backgrounds from
QCD multijet events and instrumental sources of \ETslash, which are not well-modeled by
the simulation.  We require $\ETslash>100\GeV$, near the trigger
threshold, to maximize the signal acceptance, and at least one jet
reconstructed with the PF algorithm with $\pt > 110\GeV$.
The jet must have $\abs{\eta}<2.4$ and
meet several criteria aimed at reducing instrumental noise: less than 70\% of its energy
assigned to neutral hadrons or
photons, less than 50\% of its energy associated with electrons, and
more than 20\% of its energy carried by charged hadrons. Additional
jets in the event with $\pt>30\GeV$ and $\abs{\eta}<4.5$ are allowed
provided they meet two additional criteria.  To reduce the
contribution of QCD multijets events, the difference in
azimuthal angle, $\Delta \phi$, between any two jets in the
event must be
less than 2.5 radians, and the minimum $\Delta \phi$ between the \ETslash vector
and either of the two highest-\pt jets is required to be greater than
0.5 radians.

Signal samples are simulated with \PYTHIA 6 \cite{Sjostrand:2006za} for the processes $\qqp \to \Chipm \Chiz$ and
$\qq \to \Chipm \PSGc^{\mp}_1$ in the AMSB framework.
The SUSY mass spectrum in AMSB is determined by four parameters: the
gravitino mass $m_{3/2}$, the universal scalar mass $m_0$,
the ratio of the vacuum expectation values of the Higgs field
 at the electroweak scale $\tan\beta$,
and the sign of the higgsino mass term $\sgn(\mu)$.  Of these, only
$m_{3/2}$ significantly affects the chargino mass.
We produce samples with variations of the $m_{3/2}$ parameter that
correspond to chargino masses between 100 and 600\GeV.
Supersymmetric particle mass spectra
are calculated according to the SUSY Les Houches accord \cite{SLHA} with
\ISAJET 7.80~\cite{Isajet}.
The branching fraction of the $\Chipm \to \Chiz \Pgp^\pm$ decay
is set to 100\%.
While the chargino mean proper lifetime is uniquely determined by the four
parameters above, the simulation is performed with a variety of mean proper
lifetime values ranging from 0.3 to 300 ns to expand the search beyond
the AMSB scenario.

To study the backgrounds, we use simulated samples of the
following standard model (SM) processes: $\PW + $jets, \cPqt \cPaqt,  $\cPZ\to \ell\ell$ ($\ell = \Pe,\mu,\tau$),
$\cPZ\to \cPgn \cPgn$; $\PW \PW$, $\cPZ \cPZ$, $\PW \cPZ$,
$\PW \cPgg$, and $\cPZ \cPgg$ boson pair production;
and QCD multijet and single-top-quark
production.
The $\PW + $jets and $\cPqt\cPaqt$ are generated using
\MADGRAPH5 \cite{MadGraph} with \PYTHIA6 for parton showering and
hadronization, while single top production is modeled
using \POWHEG \cite{Powheg, Powheg1, Powheg2, Powheg3} and \PYTHIA6. The $\cPZ\to\ell\ell$,
boson pair productions, and QCD multijet events are
simulated using \PYTHIA6.

All samples are simulated with CTEQ6L1 parton density functions (PDF). The full detector simulation with
\GEANTfour~\cite{GEANT4} is used to trace particles through the detector
and to model the detector response.
Additional pp interactions within a single bunch crossing (pileup) are
modelled in the simulation, and the mean number per event is
reweighted to match the number observed in data.

\section{Background characterization}
\label{sec:bkgdStudy}
In the following sections we examine the sources of both physics and
instrumental backgrounds to this search.
We consider how a disappearing track signature may be produced,
that is, a high-momentum ($\pt > 50\GeV$), isolated track without hits
in the outer layers of the tracker and with little associated energy ($ < 10\GeV$) deposited in the
calorimeters.  Various mechanisms that lead to tracks with
missing outer hits are described, and the reconstruction limitations that
impact each background category are investigated.

\subsection{Sources of missing outer hits}
\label{sec:srcMissOutHits}
A disappearing track is distinguished by missing outer hits in the tracker, \Nmissout,
those expected but not recorded after the last (farthest
from the interaction point) hit on a track.
They are calculated based on the tracker modules
traversed by the track trajectory, and they do not include modules known to
be inactive.
Standard model particles can produce tracks with missing outer hits as
the result of interactions with the tracker material.
An electron that transfers a large fraction of its energy to a
bremsstrahlung photon can change its trajectory sufficiently that
subsequent hits are not associated with the original track.
A charged hadron that interacts with a nucleus in the detector
material can undergo charge exchange, for example via
$\Pgpp + \text{n} \to\Pgpz + \Pp$, or can experience a large momentum
transfer.
In such cases, the track from the charged hadron may have no
associated hits after the nuclear interaction.

There are also several sources of missing outer hits that arise from choices made by
the default CMS tracking algorithms, which are employed in this analysis.
These allow for the possibility of missing outer hits on the tracks of
particles that traverse all of the layers of the tracker, mimicking
the signal. In a sample of simulated single-muon events we find that 11\%
of muons produce tracks that have at least one missing outer hit. This effect occurs not only with muons, but with any type of particle, and thus produces a contribution to each of the SM backgrounds.

The CMS track reconstruction algorithm identifies
many possible trajectory candidates, each constructed with
different combinations of hits.  In the case of multiple overlapping
trajectories, a
single trajectory is selected based on the number of recorded hits,
the number of expected hits not recorded, and the fit $\chi^2$.
We find that for most of
the selected trajectories with missing outer hits, there exists another candidate trajectory
without missing outer hits.

We have identified how a trajectory with missing outer
hits is chosen as the reconstructed track over a trajectory
with no missing outer hits.
The predominant mechanism is that the particle passes through
a glue joint of a double sensor module, a region of inactive material
that does not record one of the hits in between the first and last hit on
the track. Such a trajectory has no missing outer hits, but it does have
one expected hit that is not recorded.  The penalty for missing hits before
the last recorded hit is greater than for those missing after the last
hit.  As a result, the reconstructed track is instead identified as a
trajectory that stops before the layer
with the glue joint and has multiple missing outer hits.
In a smaller percentage of events, a trajectory with missing outer hits is
chosen because its $\chi^2$ is much smaller than that of a
trajectory with no missing outer hits.

\subsection{Electrons}
\label{sec:bkgdStudyElec}
We reject any tracks matched to an identified electron, but an
electron may fail to be identified if its energy is not fully
recorded by the \Ecal.
We study unidentified electrons with a \Zee tag-and-probe~\cite{WZInclusiveCrossSections}
data sample in which the tag is a well-identified electron, the probe is
an isolated track, and the invariant mass of the tag electron and
probe track is consistent with that of a $\cPZ$ boson.
From the $\eta,\phi$ distribution of probe tracks that fail to be identified as
electrons we characterize several ways that an electron's energy can be
lost.
An electron is more likely to be unidentified if it is directed
toward the overlap region between the barrel and endcap of the
\Ecal or toward the thin gaps between cylindrical sections of the barrel \Ecal.
We therefore reject tracks pointing into these regions.
An electron may also fail the identification if it is directed
towards an \Ecal channel that is inoperational or noisy, so we remove tracks
that are near any such known channels. After these vetoes, concentrations of unidentified electrons in a few regions
survive.  Thus we also veto tracks in these additional specific
regions.

\subsection{Muons}
\label{sec:bkgdStudyMu}

To reduce the background from muons, we veto tracks that are matched
to a muon meeting loose identification criteria.

We study muons that fail this identification with a \Zmumu
tag-and-probe data sample.
The probe tracks are more likely to fail
the muon identification in the region of the gap between the first two
``wheels'' of the barrel muon detector, $0.15<\abs{\eta}<0.35$; the region of
gaps between the inner and outer ``rings'' of the endcap muon disks, $1.55<\abs{\eta}<1.85$; and
in regions near a problematic muon chamber.  Tracks in these regions
are therefore excluded.

With a sample of simulated single-muon events we
investigate the signatures of muons outside these fiducial regions
that fail to be identified.
In this sample the muon reconstruction inefficiency is $6.8
\ten{-5}$.
We identify three signatures of unreconstructed muons.
One signature is a large \Ecal deposit or
a large \Hcal deposit.
In a second signature, there are reconstructed muon segments in the
muon detectors that fail to be matched to the corresponding tracker track.
The final signature has no recorded muon detector segments or
calorimeter deposits.
These signatures are consistent with a $\mu\to e\nu\cPagn$ decay in flight or a secondary
electromagnetic shower.
Lost muons that produce large calorimeter deposits are rejected, while the
contribution from those without calorimeter deposits is estimated from control samples in data. 

\subsection{Hadrons}
\label{sec:bkgdStudyTau}

Charged hadrons can produce tracks with missing outer hits as a result
of a nuclear interaction. However, tracks produced by charged hadrons in
quark/gluon jets typically
fail the requirements that the track be isolated and have little associated
calorimeter energy.
According to simulation,
the contribution from hadrons in jets in
the search sample is ten times smaller than that of the hadrons from a
single-prong hadronic tau ($\tauh$) decay.
The track from a $\tauh$ lepton decay can satisfy the criteria
of little associated calorimeter energy but
large \pt if the \pt of the hadron is mismeasured, \ie measured to
be significantly larger than the true value.
This class of background is studied using a sample of
simulated single-pion events.

In these events, the pion tracks typically have ${\approx}17$ hits.
From this original sample we produce three new samples in which
all hits associated with the track after the 5th, 6th, or 7th
innermost hit have been removed.
After repeating the reconstruction, the associated calorimeter energy
does not change with the removal of hits on the track.
However, the \pt resolution improves with the number of hits on the track, as
additional hits provide a greater lever arm to measure the track
curvature.  Thus the background from $\tauh$ decays
is largest for tracks with small numbers of hits, which motivates a
minimum number of hits requirement.

\subsection{Fake tracks}
\label{sec:fakeTrkBkgd}

Fake tracks are formed from combinations of hits that are not
produced by a single particle.
We obtain a sample of such tracks
from simulated
events that contain a track that is not matched
to any generated particle.
Most of these tracks have only three or four hits; the probability to
find a combination of hits to form a fake track decreases rapidly with
the number of hits on the track.
However, fake tracks
typically are missing many outer hits and have little associated
calorimeter energy, so they closely resemble signal tracks.

\section{Candidate track selection}
In this section, we define the {\em candidate track} criteria that are designed to
suppress the backgrounds described in the previous section
and to identify well-reconstructed, prompt tracks with large \pt.
The \candtrk sample is composed of the events that pass the
basic selection defined in Section~\ref{sec:dataset} and contain a track that meets the following criteria.

A candidate track is required to have $\pt > 50\GeV$ and $\abs{\eta} < 2.1$,
as signal tracks would typically have large \pt and are
produced centrally.
The primary vertex is chosen as the one with the largest sum $\pt^2$ of
the tracks associated to it.
The track is required to have  $\abs{d_0}< 0.02\cm$ and
$\abs{d_z}< 0.5\cm$, where $d_0$ and $d_z$ are the transverse and longitudinal impact parameters with respect
to the primary vertex. The track must be reconstructed from at
least 7 hits in the tracker.
This reduces the backgrounds associated with
poorly reconstructed tracks.

The number of
missing middle hits, \Nmissmid, is the number of hits expected but not found
between the first and last hits associated to a track. The number of
missing inner hits, \Nmissin, corresponds to lost
hits in layers of the tracker closer to the interaction point, \ie before the first hit on the track.
We require \Nmissmid = 0 and \Nmissin = 0 to ensure that the track is
not missing any hits in the pixel or strip layers before the last hit
on the track. Similarly to the calculation of missing outer hits,
the determination of missing inner and middle hits accounts
for tracker modules known to be inactive.
The relative track isolation, $(\Sigma \pt^{\Delta R<0.3} - \pt)/\pt$ must be less
than 0.05, where $\Sigma \pt^{\Delta R<0.3}$
is the scalar sum of the \pt of all other tracks within an angular distance
$\Delta R = \sqrt{\smash[b]{(\Delta \eta)^2 + (\Delta \phi)^2}} < 0.3$ of the
candidate track.  Additionally, we require that there be no
jet with $\pt > 30\GeV$ within $\Delta R<0.5$ of the track.
The above criteria select high-\pt isolated tracks.
In events with large \ETslash, the dominant SM source of high-\pt and
isolated tracks is from leptons.

We veto any tracks within $\Delta R<0.15$ of a reconstructed electron with $\pt>10\GeV$; the electron
must pass a loose identification requirement. To further reduce
the background from electrons, we veto tracks in the regions of larger
electron inefficiency described in Section~\ref{sec:bkgdStudyElec}.  These regions are
the gap between
the barrel and endcap of the \Ecal,
$1.42<\abs{\eta}<1.65$; the intermodule gaps of the \Ecal; and
all cones with aperture $\Delta R=0.05$ around inoperational or noisy \Ecal channels or
clusters of unidentified electrons in the \Zee sample.

We veto any tracks within $\Delta R<0.15$ of a muon with $\pt>10\GeV$
that passes a loose identification requirement.
We additionally reject tracks in regions of larger muon inefficiency
identified in Section~\ref{sec:bkgdStudyMu}.  These regions are
$0.15<\abs{\eta}<0.35$, $1.55<\abs{\eta}<1.85$, and
within $\Delta R<0.25$ of
any problematic muon detector.

After vetoing tracks that correspond to reconstructed electrons and muons, we face a
background from single-prong $\tauh$ decays. We veto any track within $\Delta R<0.15$ of a
reconstructed hadronic tau candidate. The reconstructed tau must have
$\pt > 30\GeV$, $\abs{\eta} < 2.3$, and satisfy a set of loose isolation
criteria.

The background contributions in the \candtrk sample, as estimated from Monte
Carlo simulations, are summarized in
Table~\ref{tab:trackGenMatchBkgd}.

\begin{table}[!htb]
  \centering
  \topcaption{The background contributions in the \candtrk sample
    estimated from the simulation by the identity of the generated particle
    matched to the candidate track.}
  \label{tab:trackGenMatchBkgd}
\begin{tabular}{lccc}
Source                            & Contribution \\
\hline
Electrons   & 15\%  \\
Muons       & 20\%  \\
Hadrons     & 60\%  \\
Fake tracks & 5\%  \\
\end{tabular}

\end{table}

\section{Disappearing track selection}
\label{sec:selDisTrk}

We define a \textit{\distrk} as a \candtrk that has the signature of missing outer hits and little associated calorimeter energy. A \distrk is first
required to have $\Nmissout \ge 3$.
Tracks from the potential signal are generally missing several outer hits, provided their
lifetime is such that they decay within the tracker volume.
To remove SM sources of tracks with missing outer hits, we additionally require
that the associated calorimeter energy \calotot of a \distrk
be less than 10\GeV, much smaller than the minimum \pt of 50\GeV. Since the decay products of the chargino are too low in momentum to be reconstructed or weakly interacting, they would not deposit significant energy in the calorimeters.
We compute \calotot as the sum of the ECAL and HCAL
clusters within $\Delta R<0.5$ of the direction
of the track.

\begin{figure}[!htb]
    \centering
        \includegraphics[width=0.48\textwidth]{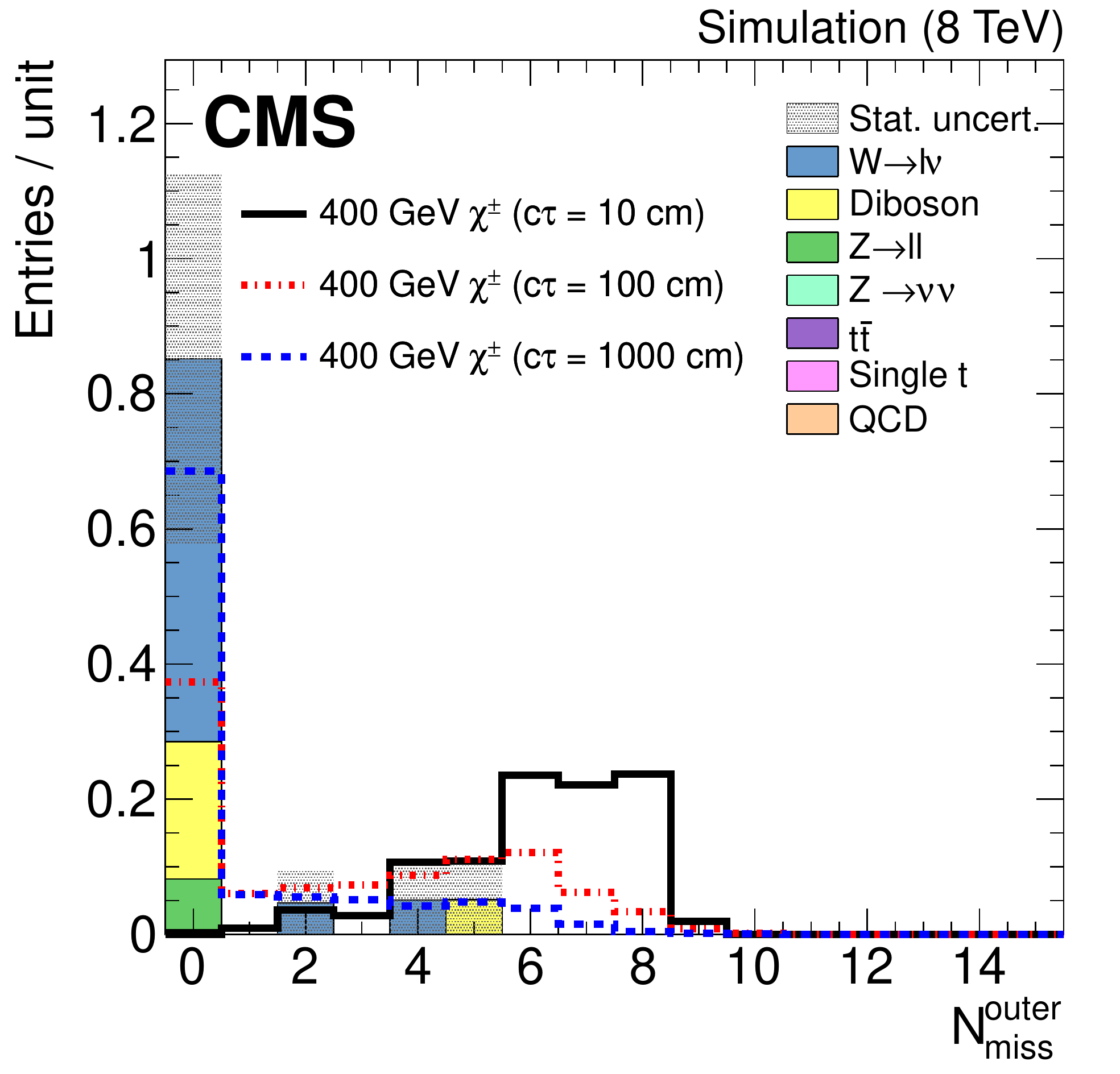}
        \includegraphics[width=0.48\textwidth]{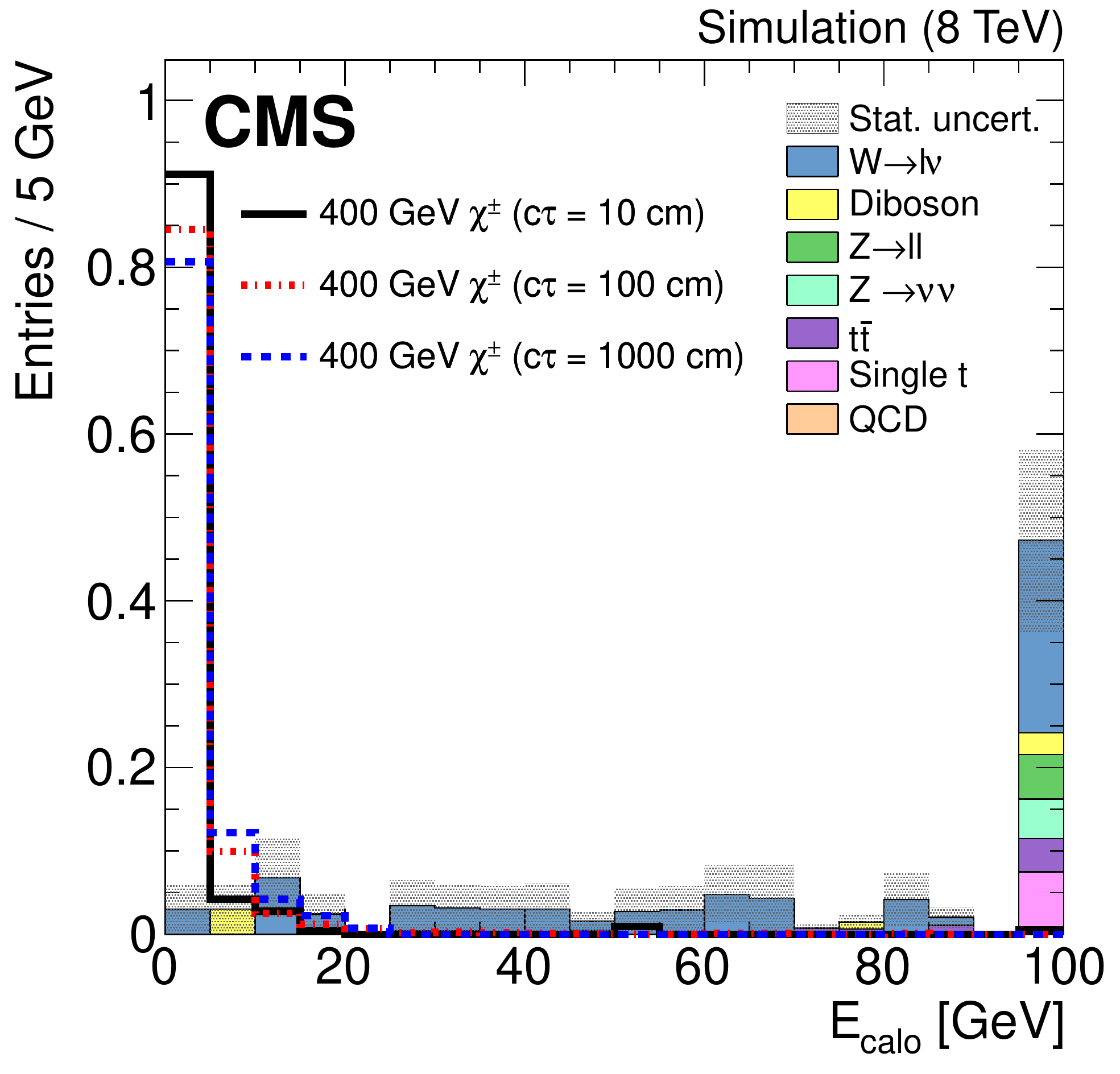}
    \caption{
      The number of missing
      outer hits (\cmsLeft) and the associated calorimeter energy (\cmsRight)
      of tracks in the search sample, before
      applying the requirement on the plotted quantity.
      The signal and the background sum distributions have both
      been normalized
      to unit area, and overflow entries
      are included in the last bin.
    }
    \label{fig:calototNmissout}

\end{figure}

The requirements placed on \calotot and \Nmissout effectively isolate
signal from background, as shown in
Fig.~\ref{fig:calototNmissout}.  Tracks produced by SM
particles generally are missing no outer hits and have large \calotot,
 while signal tracks typically have many missing
outer hits and very little \calotot.
The search sample is the subset of events in the \candtrk sample that
contain at least one \distrk.  The efficiencies to pass various stages of the selection, derived from
simulation, are given for
signal events in Table~\ref{tab:cutFlowEffSig}.

\begin{table*}[!htb]
  \centering
  \topcaption{ Cumulative efficiencies for signal events to pass various stages of the selection. }
  \label{tab:cutFlowEffSig}

  \begin{center} \begin{tabular}{lllllll}
Chargino mass\,[\GeVns{}] & 300 & 300 & 300  & 500  & 500  & 500  \\
Chargino $c\tau$\,[cm] & 10 & 100 & 1000 & 10 & 100 & 1000 \\
\hline
Trigger & 10\% & 10\% & 7.4\% & 13\% & 13\% & 10\% \\
Basic selection & 7.0\% & 6.7\% & 4.2\% & 8.9\% & 9.0\% & 6.3\% \\
High-\pt isolated track & 0.24\% & 3.6\% & 3.1\% & 0.14\% & 4.4\% & 4.9\% \\
Candidate track & 0.15\% & 2.3\% & 1.3\% & 0.10\% & 2.9\% & 2.2\% \\
Disappearing track & 0.13\% & 1.0\% & 0.27\% & 0.095\% & 1.4\% & 0.47\% \\
\end{tabular} \end{center}

\end{table*}

\section{Background estimates and associated systematic uncertainties}
For each of the background sources described in
Sections~\ref{sec:bkgdStudyElec}--\ref{sec:fakeTrkBkgd}, the
contribution in the search sample is estimated.
The SM backgrounds are estimated with a method that is based on data
and only relies on simulation to
determine the identification inefficiency.
The estimate of the fake-track background is obtained from data. 

\subsection{Standard model backgrounds}
We estimate the SM background contributions to the search sample as
$N^{i} = N^{i}_\text{ctrl}P^{i}$, where
$N^{i}_\text{ctrl}$ is the number of events in
data control samples enriched in the given background source and
$P^{i}$ is the
simulated identification inefficiency, for $i=\Pe,\Pgm,\Pgt$.
The electron-enriched control sample is selected by requiring all the
search sample criteria except for the electron veto and the
\calotot requirement.
The muon-enriched control sample is selected by requiring all the search
sample criteria except for the muon veto.  The $\tauh$-enriched control
sample is selected
by requiring all the search sample criteria except for the $\tauh$ veto
and the \calotot requirement.
The \calotot requirement is removed for the
electron and $\tauh$ control samples because it is strongly correlated
with both the electron and $\tauh$ vetoes.
The hadron background is estimated as the contribution from
$\tauh$ decays, which is its dominant component.

The identification inefficiencies $P^i$
correspond to the probability to
survive the corresponding veto criteria, \ie, the electron veto and
\calotot requirement for electrons, the muon veto for muons, and the
$\tauh$ veto and \calotot requirement for $\tau$ leptons.
We determine $P^i$, defined to be the ratio of the number of events of the given background source in the search sample to the number in the corresponding control sample, from the simulated \Wjets process.
The \Wjets process is the dominant
contribution of the control samples:  it represents 84\% of the
electron-enriched control sample, 85\% of the muon-enriched control
sample, and 75\% of the $\tauh$-enriched control sample.
Of the more than 26 million simulated \Wjets events, only one passes the search sample criteria; in that event
the disappearing track is
produced by a muon in a $\PW\to\mu\nu$ decay.
For the other simulated physics processes, no events are found in the search
sample.
Since no simulated electron or tau events survive in the search
sample, we quote limits at 68\% CL on the
electron and $\tauh$ inefficiencies.
The control sample sizes,
identification inefficiencies, and background estimates are given in
Table~\ref{tab:bkgdCalc}.

In addition to the uncertainties that result from the
finite size of the simulation samples (labeled ``statistical''), we also assess
systematic uncertainties in the simulation of $P^i$ using
tag-and-probe methods.
In \Zee, \Zmumu, and \Ztautau samples,
$P^i$ is measured as
the probability of a probe track of the given background type to
pass the corresponding veto criteria. The difference between data
and simulation is taken as the systematic uncertainty.

The probe tracks are required to pass all of the \dishtrk
criteria, with a looser requirement of $\pt > 30\GeV$, and without
the corresponding veto criteria, \ie, the electron veto and
\calotot requirement for electrons, the muon veto for muons, and the
$\tauh$ veto and \calotot requirement for taus. Additionally, to obtain
an adequate sample size, the \Ztautau probe tracks are not required to
pass the \Nmissout requirement or the isolation requirement of no
jet within $\Delta R<0.5$ of the track.
The \Zee and \Zmumu tag-and-probe samples are collected
with single-lepton triggers and require a tag
lepton ($\Pe$ or $\Pgm$) that is well-reconstructed and isolated.
The tag lepton and probe track
are required to be opposite in charge and to have an invariant mass between
80 and 100\GeV, consistent with a \Zll decay.
We measure $P^{\Pe}$ as
the fraction of \Zee probe tracks that survive the electron veto and
\calotot requirement and $P^{\Pgm}$ as the fraction of \Zmumu probe
tracks that survive the muon veto.
The \Ztautau tag-and-probe
sample is designed to identify a tag $\tau$ lepton that decays as $\Pgt\to\mu\nu\cPagn$.
This sample is collected with a single-muon trigger and
requires a well-reconstructed, isolated tag muon for which the transverse
invariant
mass of the muon \pt and \ETslash is less than 40\GeV.
The tag muon and probe track
are required to be opposite in charge and to have an invariant mass between
40 and 75\GeV, consistent with
a \Ztautau decay.
We measure $P^\Pgt$ as the fraction of probe tracks that survive the
$\tauh$ veto.  No probe tracks in the \Ztautau data survive both the $\tauh$ veto and the \calotot
requirement, so the \calotot requirement is not included in the
determination of $P^\Pgt$ for the systematic uncertainty.

For each of the tag-and-probe samples, the contamination from sources
other than the target \Zll process is estimated from
the simulation and is subtracted from both the data and simulation samples
before calculating $P^i$.
The systematic uncertainties in $P^i$ are summarized in
Table~\ref{tab:bkgdCalc}.
The systematic uncertainties in the electron and $\tauh$ estimates are
incorporated into the 68\% CL upper limit on their background contributions  according to Ref.~\cite{CousinsHighland}.

\begin{table*}[!htb]
  \centering
  \topcaption{
    The number of events in the data control samples $N^{i}_\text{ctrl}$, the
    simulated identification inefficiencies $P^i$, and the resulting estimated
    contribution in the search sample $N^i$, for each of the SM backgrounds.
    The statistical uncertainties originate from the
    limited size of the simulation samples, while the
    systematic uncertainties are derived from the differences in $P^i$
    between
    data and simulation in tag-and-probe samples.
}
  \label{tab:bkgdCalc}
\begin{tabular}{llll}
& Electrons & Muons & Taus \\
\hline
Criteria removed to   & $\Pe$ veto & $\Pgm$ veto & $\tauh$ veto \\
select control sample  & $\calotot < 10\GeV$ &            & $\calotot < 10\GeV$\\
\hline
$N^{i}_\text{ctrl}$ from data & $7785$ & $4138$  & $29$ \\
$P^{i}$ from simulation & $<6.3\times 10^{-5}$ & $1.6 ^{+3.6}_{-1.3} \times 10^{-4} $  & ${<}0.019$ \\
$N^{i} = N^{i}_\text{ctrl} P^{i} $ & ${<}0.49\stat$ & $0.64 ^{+1.47}_{-0.53}\stat$  & ${<}0.55\stat$ \\
$P^{i}$ systematic uncertainty  & 31\% & 50\% & 36\% \\
$N^{i}$  & ${<}0.50\,\text{(stat+syst)}$ & $0.64 ^{+1.47}_{-0.53}\stat\pm0.32\syst$  & $<0.57\,\text{ (stat+syst)}$ \\
\end{tabular}

\end{table*}

\subsection{Fake tracks}
\begin{figure}[!h]
\centering
        \includegraphics[width=0.48\textwidth]{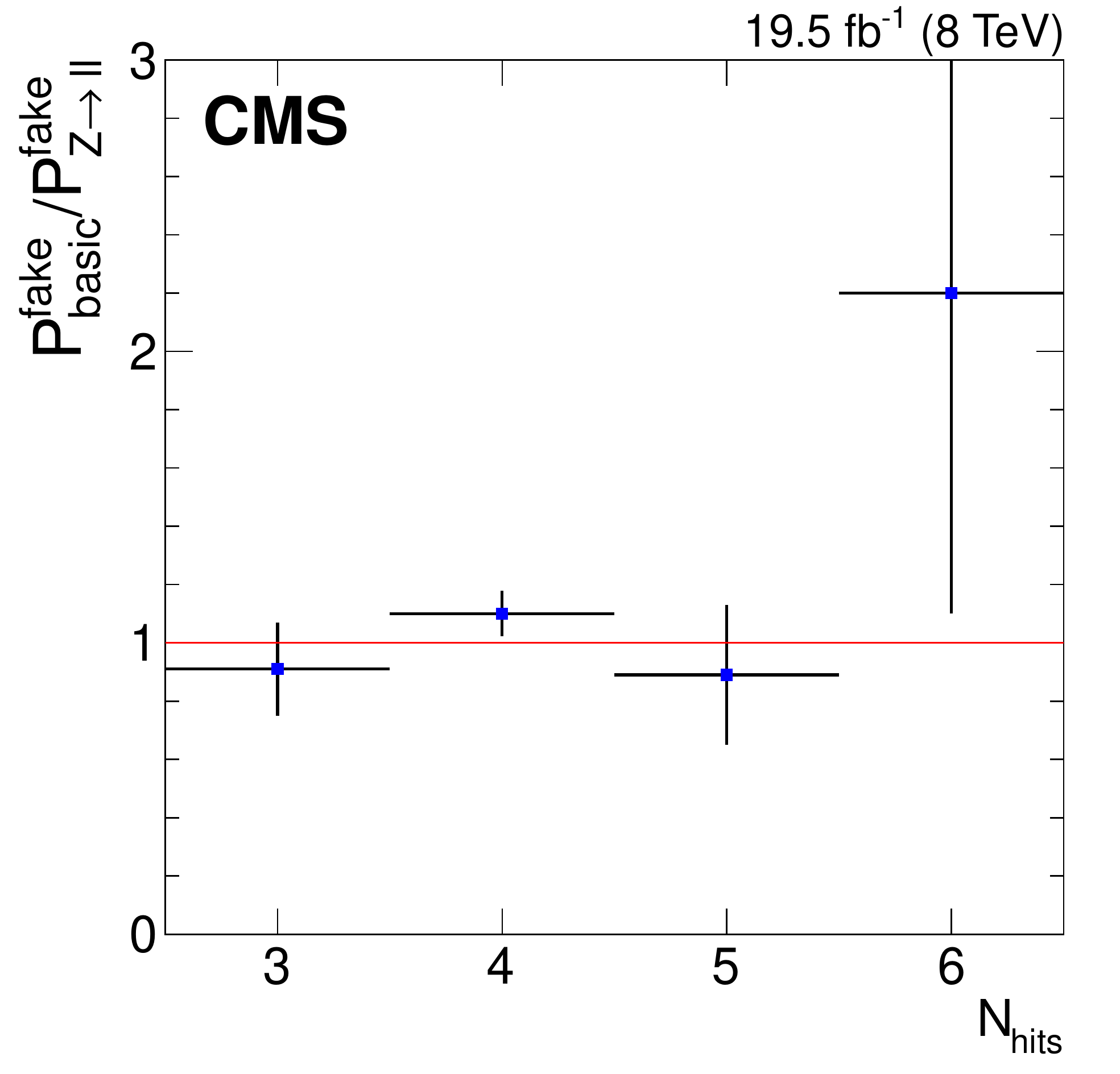}
    \caption{
      The ratio of the fake-track rates, $P^\text{fake}$, in the
      sample after the basic selection and in the \Zll control
      sample, observed in data, as a function of the number of hits on the
      candidate track.
    }
    \label{fig:fakeTrkRatios}
\end{figure}

The fake-track background is estimated as
$N^\text{fake} = N^\text{basic} P^\text{fake}$, where
$N^\text{basic} = 1.77 \ten{6}$ is the number of events in data that
pass the basic selection criteria, and
$P^\text{fake}$ is the fake-track rate determined in a \Zll ($ \ell = \Pe$ or $\mu$)
data control sample, a large sample consisting of well-understood SM processes.
In the simulation, the probability of an event to contain
a fake track that has large transverse momentum and is isolated does not depend on the
underlying physics process of the event.
The \Zll sample is collected with single-lepton triggers and
is selected by requiring two well-reconstructed, isolated leptons of
the same flavor that are
opposite in charge and have an invariant mass between 80 and 100\GeV,
consistent with a \Zll decay.
We measure $P^\text{fake}$ as the probability of
an event in the combined \Zll control sample to contain a track that passes
the \dishtrk selection.
There are two \Zll data events with an additional
track that passes the \dishtrk selection, thus $P^\text{fake}$ is
determined to be $(2.0^{+2.7}_{-1.3}) \times 10^{-7}$.
The rate of fake tracks with between 3 and 6 hits is consistent
between the sample after the basic selection and the \Zll control sample, as shown in
Fig.~\ref{fig:fakeTrkRatios}.
Fake tracks with 5 hits provide a background-enriched sample that is
independent of the search sample, in which tracks are required to have 7 or more hits.
We use the ratio of the rates
of fake tracks with 5 hits between these
two samples (including the statistical uncertainty), to assign a systematic uncertainty of
35\%.
The fake-track background estimate is
$N_\text{fake} = 0.36^{+0.47}_{-0.23}\stat\pm0.13\syst$ events.

\subsection{Background estimate validation}
The methods used to estimate the backgrounds in the search sample
are tested in three control samples:  the \candtrk sample and \calotot and \Nmissout sideband samples.
The sideband samples are depleted in signal by applying inverted
signal isolation criteria, and the size of the samples is
increased by relaxing the track \pt requirement to $\pt>30\GeV$.
In the \Nmissout
sideband sample, events must pass all criteria of the \candtrk sample,
and the candidate track must have 2 or fewer missing outer hits.
In the \calotot sideband sample, events must pass all
criteria of the \candtrk sample, and the candidate track must have more than 10\GeV
of associated calorimeter energy.
The backgrounds in each of these control samples are estimated using the methods
used to estimate the backgrounds in the search region, with the appropriate
selection criteria modified to match each sample.
The data yields and estimates in each of these samples are consistent
within the uncertainties,
as shown in Table~\ref{tab:bkgEstValidate}.
The methods of background estimation were validated in these control samples before
examining the data in the search sample.

\begin{table}[!h]
  \centering
  \topcaption{ The data yields and estimated total background in the
    \candtrk sample and the sideband samples.}
  \label{tab:bkgEstValidate}
\begin{tabular}{lccc}
Sample                                  &  Data   &  Estimate  & Data/Estimate  \\
\hline
Candidate tracks           & 59 & $49.0 \pm 5.7$ & $1.20 \pm  0.21$ \\
\calotot sideband   & 197 & $195 \pm 13$ & $1.01 \pm  0.10$ \\
\Nmissout sideband & 112 & $103 \pm 9$ & $1.09 \pm  0.14$ \\
\end{tabular} \\
\end{table}

\section{Additional systematic uncertainties}
\label{sec:syst}
In addition to the systematic uncertainties in the background
estimates described previously, there are systematic uncertainties associated with
the integrated luminosity and the signal efficiency.

The integrated luminosity of the 8\TeV \pp collision data is measured
with a pixel cluster counting
method, for which the uncertainty is 2.6\%~\cite{CMS-PAS-LUM-13-001}.

The uncertainty associated with the simulation of jet radiation is assessed by
comparing the  recoil of muon pairs from ISR jets in data with a sample of \PYTHIA  simulated  $\Zmumu$+jets events. The dimuon spectra ratio of data to simulation is used to weight the signal
events, and the corresponding selection criteria efficiency is compared
to the nominal efficiency.
The uncertainty is 3--11\%.

We assess
uncertainties due to the jet energy scale and resolution
from the effect of varying up and down by one standard
deviation the jet energy corrections and jet energy resolution
smearing parameters~\cite{JESPaper}.
The selection efficiency changes by 0--7\% from the variations in the
jet energy corrections and jet
energy resolution.

We assess the PDF uncertainty by evaluating the
envelope of uncertainties of the CTEQ6.6, MSTW08,
and NNPDF2.0 PDF sets, according to the PDF4LHC
recommendation~\cite{pdf2,Alekhin:2011sk}.
The resultant acceptance uncertainties are 1--10\%.

The uncertainty associated with the trigger efficiency is assessed with a
sample of $\PW\to \mu \nu$ events.
We compare the trigger efficiency in data and simulation
as a function of \ETslash reconstructed after excluding muons, as the trigger
efficiency is
similar for $\PW\to \mu \nu$ and signal events. We select $\PW\to \mu \nu$ events by applying the basic selection criteria
excluding the \ETslash requirement. We also apply the  candidate track criteria
excluding the muon veto.
The ratio of the trigger efficiency in data and simulation
is used to weight the signal events.  The resultant
change in the selection efficiency is 1--8\%.

The uncertainty associated with the modeling of \Nmissout is
assessed by varying the \Nmissout distribution of the simulated signal samples
by the disagreement between data and simulation in the
 \Nmissout distribution in a control sample of muon tracks.
Since muons are predominantly affected by the algorithmic sources of missing outer hits
described in Section~\ref{sec:srcMissOutHits}, they illustrate
how well the \Nmissout distribution is modeled in simulation.
The consequent change in signal efficiencies is found to be 0--7\%.

The uncertainties associated with missing inner and
middle hits are assessed as the difference between data and simulation in the
efficiency of the requirements of zero missing inner or middle hits in
a control sample of muons.
A sample of muons is used because they produce tracks
that rarely have missing inner or middle hits, as would be the case
for signal.
These uncertainties are 3\% for missing inner hits and
2\% for missing middle hits.

The systematic uncertainty associated with the simulation of
\calotot is assessed as the difference between data and simulation in the
efficiency of the \calototCut requirement, in a control sample of fake
tracks with exactly 4 hits.  This sample is used
because such tracks have very little associated
calorimeter energy, as would be the case for signal tracks.
The uncertainty is 6\%.

The uncertainty associated with the modeling of the number of
pileup interactions per bunch crossing
is assessed by weighting the signal events to match target pileup distributions in
which the numbers of inelastic interactions are shifted up and down by
the uncertainty.  The consequent variation in the signal efficiency is 0--2\%.

The uncertainty in the track reconstruction efficiency is assessed
with a tag-and-probe study ~\cite{CMS-PAS-TRK-10-002}.  The track
reconstruction efficiency is measured for probe muons, which are reconstructed
using information from the muon system only.
We take the uncertainty to be the largest difference between data and
simulation among several pseudorapidity ranges, observed to be 2\%.

The systematic uncertainties in the signal efficiency
for samples of charginos with $c\tau$ in the range of maximum
sensitivity, 10--1000\cm, and all simulated masses,
are summarized in Table~\ref{tab:sigSyst}.

\begin{table}[!h]
  \centering
  \topcaption{Signal efficiency systematic uncertainties, for
    charginos with masses in the range 100--600\GeV and $c\tau$
 of 10--1000\cm. }
  \label{tab:sigSyst}
\begin{tabular}{ll}
Jet radiation (ISR)                 & 3--11\% \\
Jet energy scale / resolution       & 0--7\% \\
PDF                                 & 1--10\% \\
Trigger efficiency                  & 1--8\% \\
\Nmissout modeling                  & 0--7\% \\
\Nmissin, \Nmissmid modeling        & 2--3\% \\
\calotot modeling                   & 6\% \\
Pileup                              & 0--2\% \\
Track reconstruction efficiency     & 2\% \\
\hline
Total                               & 9--22\% \\
\end{tabular}
\end{table}

\section{Results}
\label{sec:results}

Two data events are observed in the search sample, which is consistent
with the expected background. The numbers of expected events from
background sources compared with data in the search
sample are shown in Table~\ref{tab:bkgEstSumm}.
From these results, upper limits at 95\% CL on the total production cross
section of direct electroweak chargino-chargino
and chargino-neutralino production are calculated
for various chargino masses and mean proper lifetimes.
The next-to-leading-order cross sections for these processes, and
their uncertainties, are taken
from Refs.~\cite{CharginoProduction1999, SusyProduction}. The limits are calculated with the CL$_\mathrm{S}$
technique~\cite{Read:2002hq, Junk:1999kv},
using the
LHC-type CL$_\mathrm{S}$ method~\cite{CMS_ATLAS_HiggsCombination}.
This method uses a test statistic based on a profile likelihood
ratio~\cite{asymptoticCLs}
and treats nuisance parameters in a frequentist context.
Nuisance parameters for the systematic uncertainties in the integrated
luminosity and in the signal efficiency are constrained with log-normal
distributions.
There are two types of nuisance parameters for the uncertainties in
the background estimates, and they are
specified separately for each of the four background contributions.
Those that result from the limited size of a sample are
constrained with gamma distributions, while
those that are associated with the relative disagreement between data and
simulation in a control region have log-normal constraints.
The mean and standard deviation of the
distribution of pseudo-data generated under the background-only hypothesis
provide an estimate of the total background contribution to the search sample of $1.4 \pm1.2$ events.

\begin{table}[!htb]
  \centering
  \topcaption{ The expected background from all sources compared with data in the search
    sample. }
  \label{tab:bkgEstSumm}
\begin{tabular}{llll}
Event source    &  \multicolumn{2}{c}{Yield}                  \\
\hline
Electrons      & $ {<}0.49\stat$  & ${<}0.50\,\text{(stat+syst)}$ \\
Muons          & \multicolumn{2}{c}{$0.64^{+1.47}_{-0.53}\stat\pm 0.32\syst$ }  \\
Taus           & $ {<}0.55\stat$ & ${<}0.57\,\text{(stat+syst)} $ \\
Fake tracks    & \multicolumn{2}{c}{$0.36^{+0.47}_{-0.23}\stat\pm 0.13\syst$ }  \\
\hline
Data           & \multicolumn{2}{c}{ $ 2 $ }  \\
\end{tabular}

\end{table}

\begin{figure*}[!htbp]
\centering
        \includegraphics[width=0.48\textwidth]{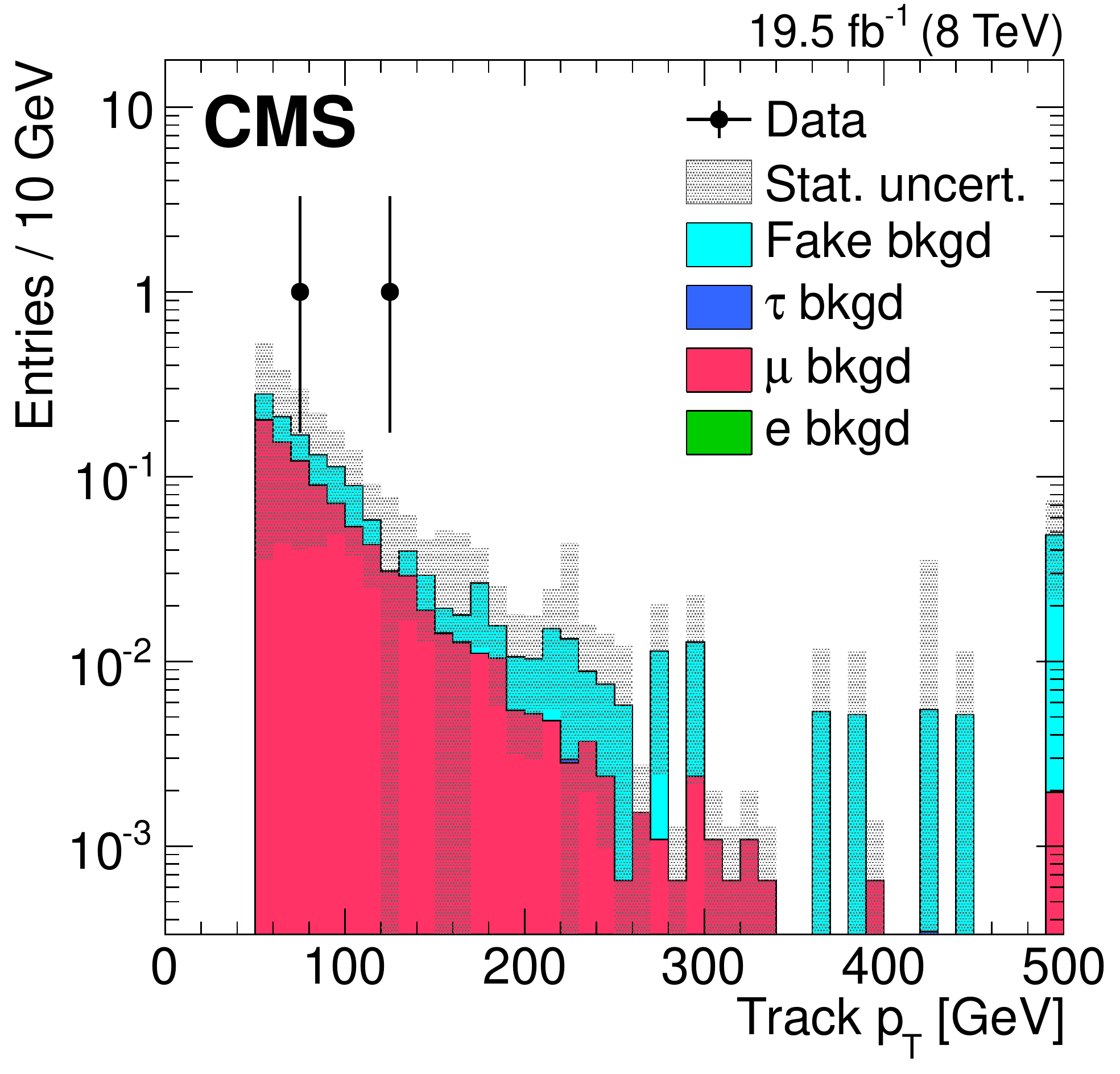}
        \includegraphics[width=0.48\textwidth]{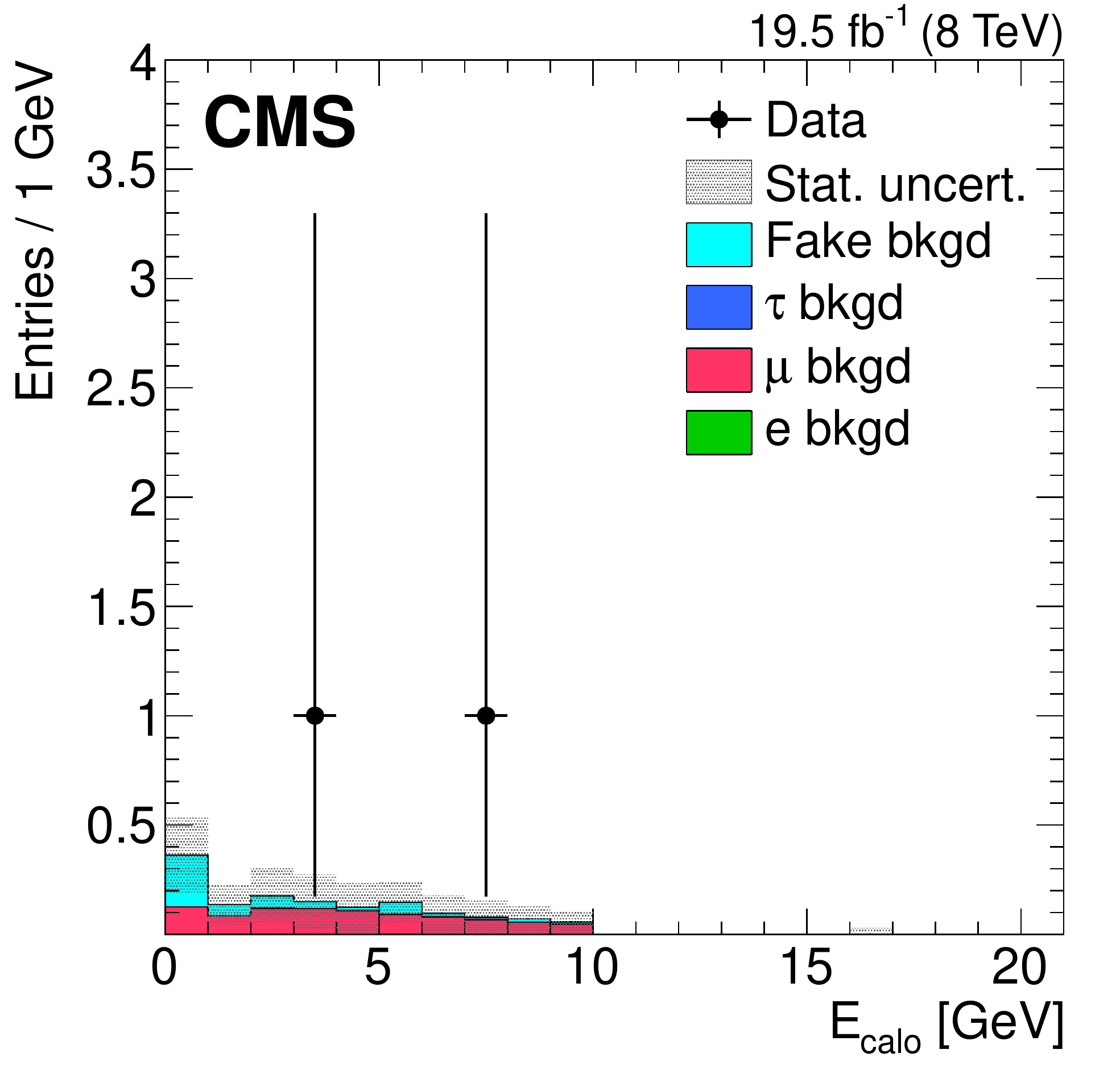}
        \includegraphics[width=0.48\textwidth]{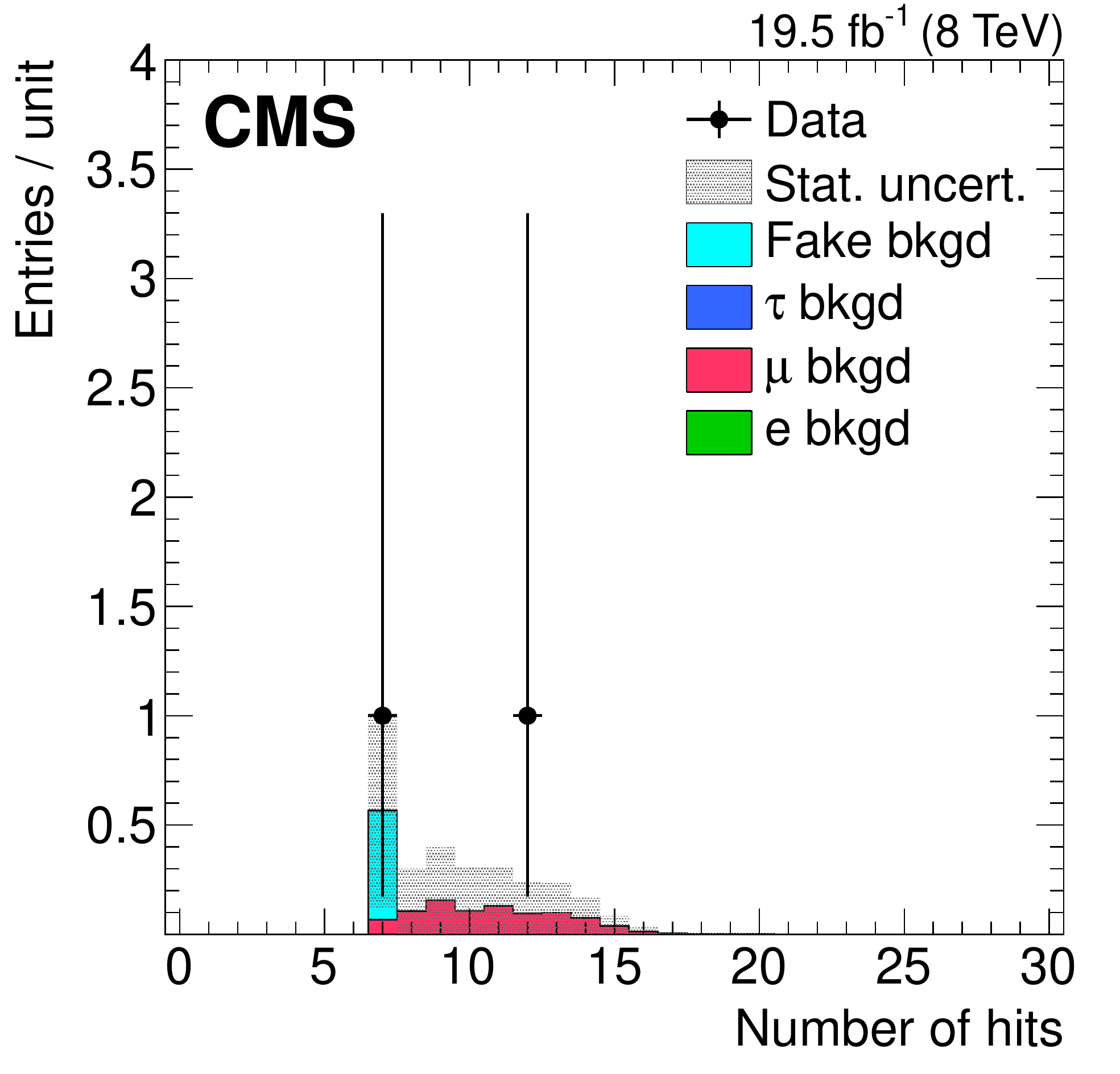}
        \includegraphics[width=0.48\textwidth]{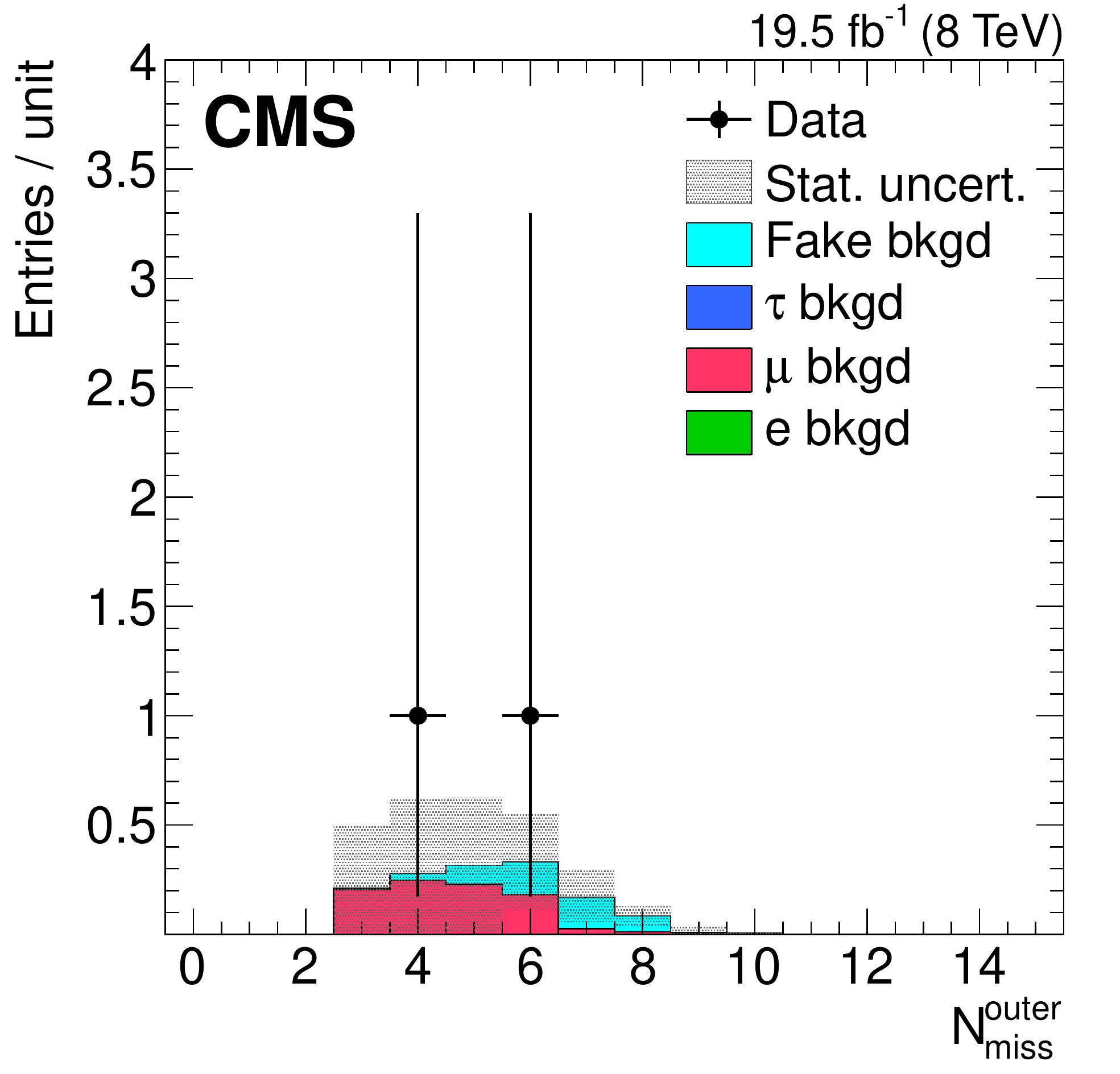}
    \caption{
      Distributions of the disappearing tracks in the search sample.
      The estimated backgrounds are normalized to have the relative contributions
      of Table~\ref{tab:bkgEstSumm} and a total equal to the
      mean of the background-only pseudo-data (1.4 events).
      Histograms for the electron and tau backgrounds are not visible
      because the central value of their estimated contribution is zero.
    }
    \label{fig:distDataBkgd_FullSel}
\end{figure*}

The distributions of the \pt, number of hits, \calotot, and
\Nmissout of the disappearing tracks in the search region are shown
for the observed events and the
estimated backgrounds in Fig.~\ref{fig:distDataBkgd_FullSel}. The
shapes of the electron, muon, and tau
background distributions are obtained from the data control samples
enriched in the given background.
The fake track distribution shapes are taken from the \Zll
control sample, using fake tracks with 5 hits, except for the
plot of the number of hits, for which fake tracks with 7 or more hits are
used.
The background normalizations have the relative contributions
of Table~\ref{tab:bkgEstSumm} and a total equal to 1.4 events, the
mean of the background-only pseudo-data.
No significant discrepancy
between the data and estimated background is found.

In contrast to a slowly moving chargino, which is expected
to have a large average ionization energy loss, the energy loss of the two
disappearing tracks in the search sample is compatible with that of
minimum-ionizing SM particles, ${\approx}3\MeV/$cm.

\begin{figure}[!htb]
 \centering
     \includegraphics[width=\cmsFigWidth]{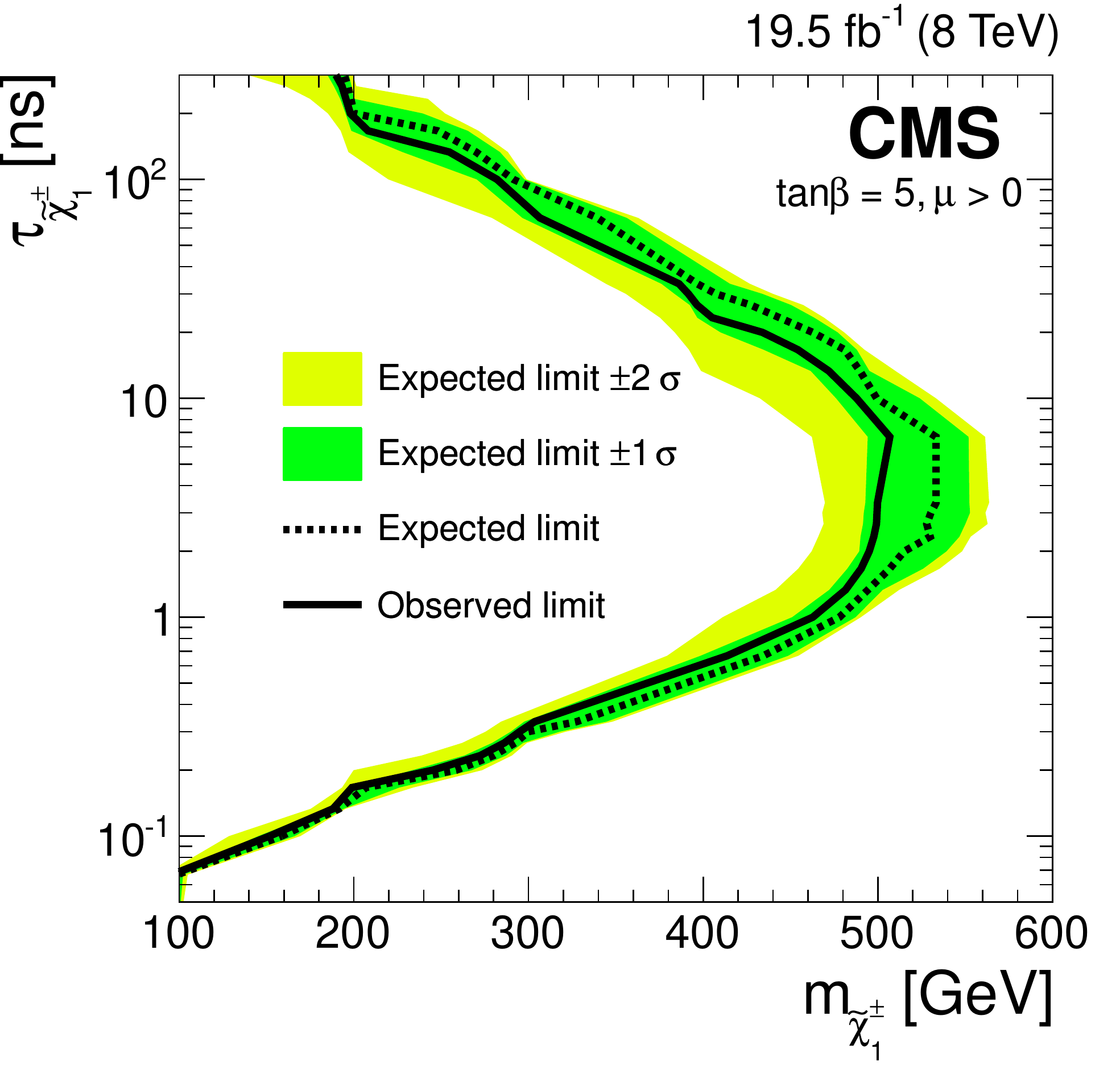}
    \caption{The expected and observed constraints on the chargino mean proper lifetime and mass.  The
      region to the left of the curve is excluded at 95\% CL.
    }
    \label{fig:limits}
\end{figure}

The expected and observed constraints on the allowed chargino mean proper lifetime and mass are presented in
Fig.~\ref{fig:limits}.
The maximum sensitivity is for charginos with a mean proper lifetime of
7\unit{ns}, for which masses less than 505\GeV are excluded at 95\% CL.

\begin{figure}[!htb]
\centering
      \includegraphics[width=0.48\textwidth]{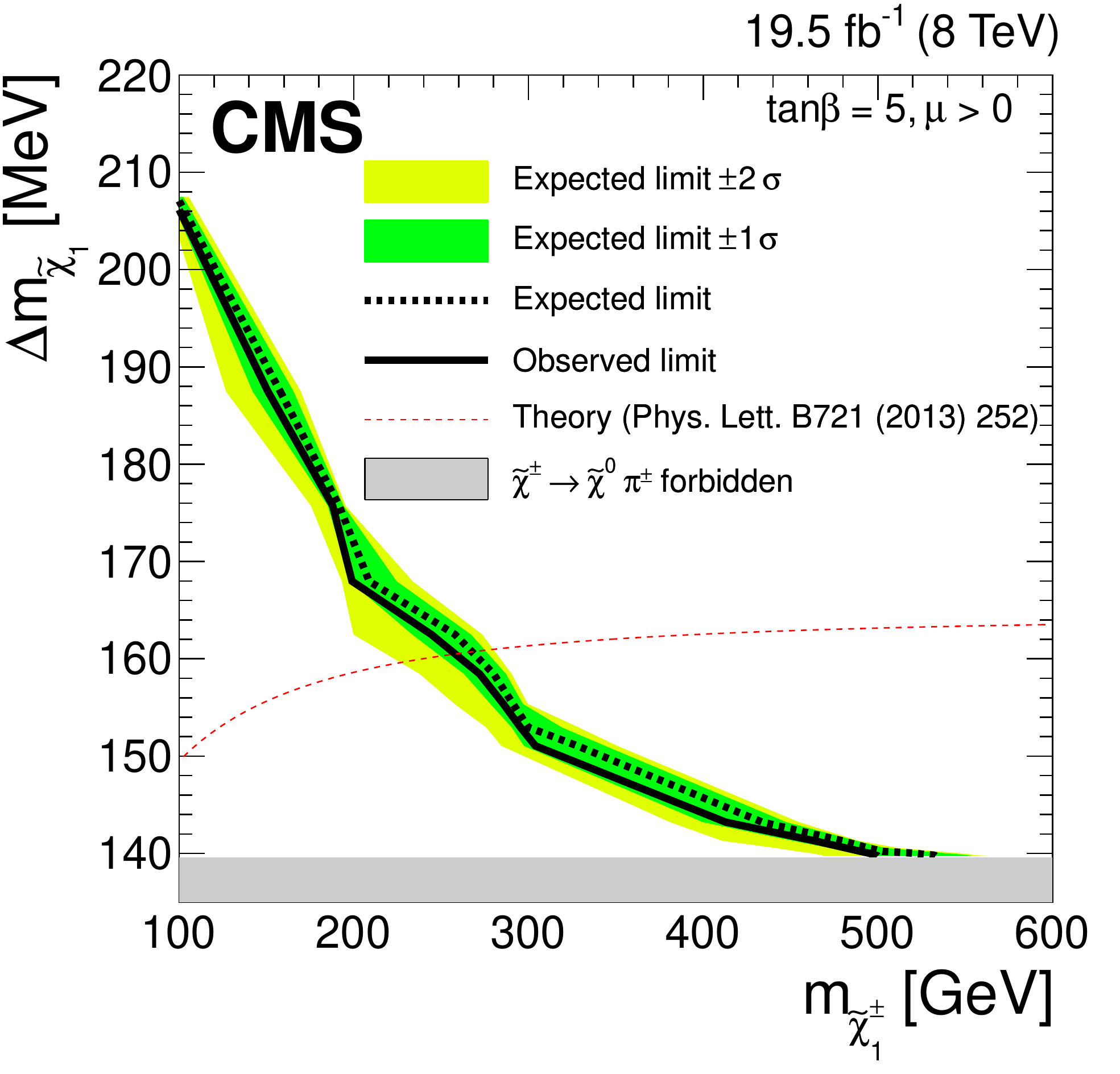}
    \caption{The expected and observed constraints on the chargino mass and the mass
      splitting between the chargino and neutralino, $\Delta m_{\PSGc_1}$, in the AMSB model.  The
      prediction for $\Delta m_{\PSGc_1}$ from Ref.~\cite{massSplittingCurve} is
      also indicated.
    }
    \label{fig:massSplit}
\end{figure}

In Fig.~\ref{fig:massSplit}, we show the expected and observed constraints on the
mass of the chargino and the mass difference between the chargino and
neutralino, $\Delta m_{\PSGc_1} = m_{\PSGc^\pm_1} - m_{\PSGc^0_1}$,
in the minimal AMSB model. The limits on $\tau_{\PSGc^\pm_1}$ are converted into limits on $\Delta m_{\PSGc_1}$
according to Ref.~\cite{equation, equation2}. The two-loop level calculation of $\Delta m_{\PSGc_1}$
for wino-like lightest chargino and neutralino states~\cite{massSplittingCurve}
is also indicated. In the AMSB model, we exclude
charginos with mass less than 260\GeV, corresponding to a chargino mean proper
lifetime of 0.2\unit{ns} and $\Delta m_{\PSGc_1} = 160\MeV$.

\begin{figure}[!htb]
\centering
      \includegraphics[width=\cmsFigWidth]{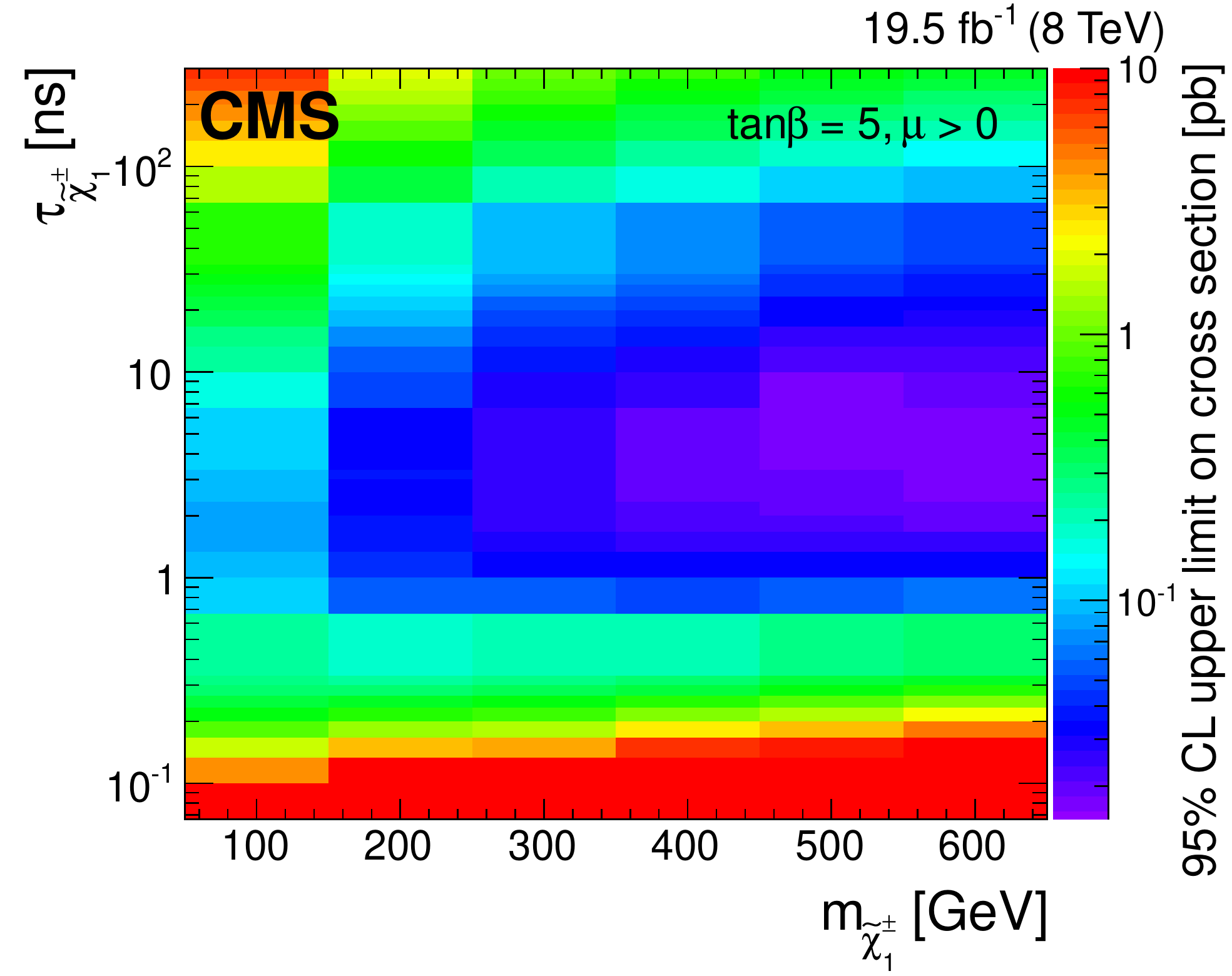}
    \caption{The observed upper limit (in pb) on the total cross
      section of $\qqp \to \Chipm \Chiz$
       and $\qq \to \Chipm \PSGc^{\mp}_1$ processes as a function
      of chargino mass and mean proper lifetime.
      The simulated chargino mass used to obtain the limits
      corresponds to the center of each bin.
    }
    \label{fig:3DUpperLimit}
\end{figure}

In Fig.~\ref{fig:3DUpperLimit}, we show the observed upper limit on the total cross
section of the $\qqp \to \Chipm \Chiz$ plus
$\qq \to \Chipm \PSGc^{\mp}_1$ processes in terms of chargino mass and mean proper lifetime.
A model-independent interpretation of the results is provided in \ifthenelse{\boolean{cms@external}}{}{Appendix} \ref{sec:appendix}.

\section{Summary}
\label{sec:conclusion}

A search has been presented for long-lived charged particles that decay within the
CMS detector and produce the signature of a disappearing track.
In a sample of proton-proton data recorded at a collision energy of $\sqrt{s}=8\TeV$ and corresponding to
an integrated luminosity of 19.5\fbinv, two events are observed in the search sample.
Thus, no significant excess above the estimated background of $1.4 \pm1.2$
events is observed
and constraints are placed on the chargino mass, mean proper lifetime, and mass
splitting.
Direct electroweak production of charginos with a mean proper lifetime of 7\unit{ns} and
a mass less than 505\GeV is excluded at 95\% confidence level.
In the AMSB model, charginos with masses less than
260\GeV, corresponding to a mean proper lifetime of  0.2\unit{ns} and chargino-neutralino
mass splitting of $160\MeV$, are excluded at 95\% confidence level.
These constraints corroborate those set by the ATLAS
Collaboration~\cite{ATLASDisapp2}.

\begin{acknowledgments}
We congratulate our colleagues in the CERN accelerator departments for
the excellent performance of the LHC and thank the technical and
administrative staffs at CERN and at other CMS institutes for their
contributions to the success of the CMS effort. In addition, we
gratefully acknowledge the computing centers and personnel of the
Worldwide LHC Computing Grid for delivering so effectively the
computing infrastructure essential to our analyses. Finally, we
acknowledge the enduring support for the construction and operation of
the LHC and the CMS detector provided by the following funding
agencies: BMWFW and FWF (Austria); FNRS and FWO (Belgium); CNPq,
CAPES, FAPERJ, and FAPESP (Brazil); MES (Bulgaria); CERN; CAS, MoST,
and NSFC (China); COLCIENCIAS (Colombia); MSES and CSF (Croatia); RPF
(Cyprus); MoER, ERC IUT, and ERDF (Estonia); Academy of Finland, MEC,
and HIP (Finland); CEA and CNRS/IN2P3 (France); BMBF, DFG, and HGF
(Germany); GSRT (Greece); OTKA and NIH (Hungary); DAE and DST (India);
IPM (Iran); SFI (Ireland); INFN (Italy); NRF and WCU (Republic of
Korea); LAS (Lithuania); MOE and UM (Malaysia); CINVESTAV, CONACYT,
SEP, and UASLP-FAI (Mexico); MBIE (New Zealand); PAEC (Pakistan); MSHE,
and NSC (Poland); FCT (Portugal); JINR (Dubna); MON, RosAtom, RAS and
RFBR (Russia); MESTD (Serbia); SEIDI and CPAN (Spain); Swiss Funding
Agencies (Switzerland); MST (Taipei); ThEPCenter, IPST, STAR, and NSTDA
(Thailand); TUBITAK and TAEK (Turkey); NASU and SFFR (Ukraine); STFC
(United Kingdom); DOE and NSF (USA).
Individuals have received support from the Marie-Curie program and
the European Research Council and EPLANET (European Union); the
Leventis Foundation; the A. P. Sloan Foundation; the Alexander von
Humboldt Foundation; the Belgian Federal Science Policy Office; the
Fonds pour la Formation \`a la Recherche dans l'Industrie et dans
l'Agriculture (FRIA-Belgium); the Agentschap voor Innovatie door
Wetenschap en Technologie (IWT-Belgium); the Ministry of Education,
Youth and Sports (MEYS) of the Czech Republic; the Council of Science
and Industrial Research, India; the HOMING PLUS program of
Foundation for Polish Science, cofinanced from European Union,
Regional Development Fund; the Compagnia di San Paolo (Torino); the
Consorzio per la Fisica (Trieste); MIUR project 20108T4XTM (Italy);
the Thalis and Aristeia programs cofinanced by EU-ESF and the Greek
NSRF; and the National Priorities Research Program by Qatar National
Research Fund.

Individuals have received support from the Marie-Curie program and the European Research Council
and EPLANET (European Union); the Leventis Foundation; the A. P. Sloan Foundation; the Alexander
von Humboldt Foundation; the Belgian Federal Science Policy Office; the Fonds pour la Formation \`a la
Recherche dans l'Industrie et dans l'Agriculture (FRIA-Belgium); the Agentschap voor Innovatie door
Wetenschap en Technologie (IWT-Belgium); the Ministry of Education, Youth and Sports (MEYS) of the
Czech Republic; the Council of Science and Industrial Research, India; the HOMING PLUS program
 of Foundation for Polish Science, cofinanced from European Union, Regional Development Fund; the
 Compagnia di San Paolo (Torino); the Consorzio per la Fisica (Trieste); MIUR project 20108T4XTM (Italy);
the Thalis and Aristeia programmes cofinanced by EU-ESF and the Greek NSRF; and the National Priorities
Research Program by Qatar National Research Fund.
\end{acknowledgments}
\bibliography{auto_generated}   
\appendix
\section{Model-independent interpretation}
\label{sec:appendix}
To allow the interpretation of the results of this search in the context of other
new physics models, the signal efficiency is parameterized in terms of
the four-momenta and decay positions of the generated BSM particles.  This allows
the signal efficiency to be approximated without performing a full
simulation of the CMS detector.
In this approximation, the signal efficiency is factorized as
$\epsilon = \epsilon_\mathrm{b}\epsilon_\mathrm{t}$, where
$\epsilon_\mathrm{b}$ is the probability of an event to pass the basic
selection and $\epsilon_\mathrm{t}$ is the probability for that event
to contain at least one disappearing track.
The efficiency to pass the basic selection $\epsilon_\mathrm{b}$
depends mostly on the \pt
of the BSM system, which is approximately equal to \ETslash.  The efficiency
of the basic selection as a function of the \pt of the chargino-chargino or
chargino-neutralino system \pt($\PSGc\PSGc$) is shown in
Table~\ref{tab:modelIndepBasic}.
To calculate the probability $\epsilon_\mathrm{t}$ that an event contains a disappearing
track, it is necessary to first identify charged particles that pass the following
track preselection criteria:
$\pt > 50$\GeV, $\abs{\eta} < 2.2$, and a decay position within the
tracker volume, \ie, with a longitudinal distance to the interaction
point of less than 280\cm and
a transverse decay distance in the laboratory frame $L_{xy}$ of less than 110\cm.
For long-lived charged particles that meet the track preselection
criteria, the efficiency to pass the \dishtrk selection depends mostly
on $L_{xy}$, as given in Table~\ref{tab:modelIndepDisTrk}.
Each of the long-lived BSM particles that pass the
preselection should be considered, weighted by its \dishtrk efficiency
from Table~\ref{tab:modelIndepDisTrk}, to determine whether the event
contains at least one disappearing track.
This parameterization of the efficiency is valid under the
assumptions that the long-lived BSM particles are isolated and
that their decay products deposit little or no energy in the calorimeters.
For the benchmark signal samples used in this analysis, the
efficiency approximation agrees with the full simulation
efficiencies given in Table~\ref{tab:cutFlowEffSig} within 10\% for
charginos with \ctau between 10 and 1000\cm.
The expected number of signal events $N$ for a new physics process is the product
 of the signal
efficiency $\epsilon$, the cross section $\sigma$, and the
integrated luminosity $L$, $N = \epsilon \sigma L$.
By comparing such a
prediction with the estimated background of $1.4 \pm 1.2$ events and
the observation of two
events in this search, constraints on other models can be set.

\begin{table}[!htb]
  \centering
  \topcaption{ Efficiency of an event to pass the basic
    selection. Uncertainties are statistical only.  }
  \label{tab:modelIndepBasic}
\begin{tabular}{lc}
$\pt(\tilde{\chi}\tilde{\chi})$\,[\GeVns{}]    &  Basic selection efficiency (\%) \\
\hline
   ${<}100$ & 0.0 $\pm$ 0.0 \\
100--125  & 13.1 $\pm$ 0.3 \\
125--150  & 44.1 $\pm$ 0.8 \\
150--175  & 65.3 $\pm$ 1.2 \\
175--200  & 75.7 $\pm$ 1.5 \\
200--225  & 79.5 $\pm$ 1.9 \\
   ${>}225$ & 85.5 $\pm$ 1.1 \\
\end{tabular}

\end{table}

\begin{table}[!htb]
  \centering
  \topcaption{ Efficiency of a track to pass the \dishtrk
    selection after passing the preselection as a function of the transverse
    decay distance in the laboratory frame, $L_{xy}$ . Uncertainties are
    statistical only. }
  \label{tab:modelIndepDisTrk}
\begin{tabular}{lc}
$L_{xy}$\,[cm]    &  Disappearing track efficiency (\%) \\
\hline
   ${<}30$ & 0.0 $\pm$ 0.2 \\
30--40  & 26.0 $\pm$ 1.0 \\
40--50  & 44.2 $\pm$ 1.6 \\
50--70  & 50.8 $\pm$ 1.4 \\
70--80  & 45.5 $\pm$ 2.1 \\
80--90  & 25.5 $\pm$ 1.6 \\
90--110  & 3.1 $\pm$ 0.4 \\
   ${>}110$ & 0.0 $\pm$ 0.0 \\
\end{tabular}

\end{table}

\cleardoublepage \section{The CMS Collaboration \label{app:collab}}\begin{sloppypar}\hyphenpenalty=5000\widowpenalty=500\clubpenalty=5000\textbf{Yerevan Physics Institute,  Yerevan,  Armenia}\\*[0pt]
V.~Khachatryan, A.M.~Sirunyan, A.~Tumasyan
\vskip\cmsinstskip
\textbf{Institut f\"{u}r Hochenergiephysik der OeAW,  Wien,  Austria}\\*[0pt]
W.~Adam, T.~Bergauer, M.~Dragicevic, J.~Er\"{o}, M.~Friedl, R.~Fr\"{u}hwirth\cmsAuthorMark{1}, V.M.~Ghete, C.~Hartl, N.~H\"{o}rmann, J.~Hrubec, M.~Jeitler\cmsAuthorMark{1}, W.~Kiesenhofer, V.~Kn\"{u}nz, M.~Krammer\cmsAuthorMark{1}, I.~Kr\"{a}tschmer, D.~Liko, I.~Mikulec, D.~Rabady\cmsAuthorMark{2}, B.~Rahbaran, H.~Rohringer, R.~Sch\"{o}fbeck, J.~Strauss, W.~Treberer-Treberspurg, W.~Waltenberger, C.-E.~Wulz\cmsAuthorMark{1}
% ...
\vskip\cmsinstskip
% ...
1:~~Also at Vienna University of Technology, Vienna, Austria\\
2:~~Also at CERN, European Organization for Nuclear Research, Geneva, Switzerland\\
% ...

\end{sloppypar}
\end{document}